\documentclass[12pt,a4paper]{article}
\usepackage[T1]{fontenc}
\usepackage{lmodern}
\usepackage[margin=25mm]{geometry}
\usepackage[numbers,sort&compress,square]{natbib}
\usepackage{setspace}
\usepackage{placeins}
\usepackage{float}
\usepackage{needspace}
\usepackage{booktabs}
\usepackage{amsmath,amssymb,graphicx}
\usepackage[colorlinks=true,linkcolor=blue,citecolor=blue,urlcolor=blue]{hyperref}
\input{glyphtounicode}
\newcommand{\X}{X(3872)}
\newcommand{\Ps}{\psi(2S)}
\newcommand{\Jp}{J/\psi}
\newcommand{\pt}{p_{\mathrm T}}
\newcommand{\ee}{\mathrm e}
\hypersetup{pdftitle={Causal-Horizon scaling of prompt X(3872) formation lengths},pdfauthor={Yi Yang}}
\begin{document}
\title{Causal-Horizon scaling of prompt X(3872) formation lengths}
\author{Yi Yang\thanks{Contact author: \texttt{yiyang429@as.edu.tw}}\\[0.6em]
 {\small Institute of Physics, Academia Sinica, Taipei 11529, Taiwan}\\
 {\small Department of Physics, National Cheng Kung University, Tainan 70101, Taiwan}}
\date{October 30, 2026}

\begingroup
\setstretch{1}
\maketitle
\endgroup

\begin{abstract}
A causal-horizon ensemble relates changes in prompt production
to spatial correlations involved in forming $\X$.
Each horizon represents the spatial coherence range of an
early-field pulse. For weak phase fluctuations and overlap widths small
compared with the mean horizon, ensemble averaging makes the fractional
loss of formation probability proportional to the Gaussian source-state
overlap width. Extending this dependence as an effective law at finite
attenuation, we compare three states to eliminate the common field
coefficient. A common-slope fit to matched LHCb prompt $\X/\Ps$ and
$\Ps/\Jp$ ratios at 13 TeV gives conditional formation scales.
With a common Gaussian source, common field coupling, fixed relative
production baselines and a direct-$\Jp$ reference, the central fit requires
a source root-mean-square width above about 0.60 fm and an $\X$ effective
Gaussian RMS separation of at least about 1.25 fm.
The same assumptions require at least a factor of about 4.5 increase in
the $\X$ formation probability relative to its no-field value between
4--6 and 12--14 GeV/$c$ in $2<y<3$.
Calculations with finite feed-down and a nodal reference assess the length
calibration. A matched three-state activity measurement tests the same
spatial slope without an additional activity-dependent parameter.
\end{abstract}

\FloatBarrier
\section{Introduction}
\label{sec:intro}
The internal structure of $\X$ remains unsettled, although its
$1^{++}$ quantum numbers are established~\cite{Belle,Quantum,XYZreview}.
Its mass and decay distributions constrain near-threshold dynamics,
whereas prompt production tests how the correlations that eventually
form the state are prepared. These processes need not sample the
same part of a hadronic wave function~\cite{Factorization,Production,ProductionScales}.
We ask over what spatial extent the precursor must be correlated
for an early field to modify its probability of forming $\X$.
Throughout this paper, a ``formation length'' means an internal spatial
extent, rather than the distance traveled before binding.

The causal-horizon approach was developed to describe an early
contribution to quarkonium suppression in heavy-ion
collisions~\cite{Yang}. In that picture, strong early color
fields introduce a fluctuating causal length, and the formation
probability depends on the spatial extent of the state relative
to that length. Using potential-model bottomonium radii, a common
horizon scale describes the observed suppression hierarchy and
the main LHC centrality dependence without adjusting the early
factor separately for each state. Its local, scalar contribution
is also consistent with the small measured bottomonium elliptic
flow. The measured excited-to-ground-state double ratio requires
an additional contribution from later evolution.
These results motivate asking whether the response to an early
field can also constrain the correlations involved in $\X$ formation.

Here we consider a source prepared by the collision, an early
field that acts on its spatial correlations, and a final projection
onto a hadron. The complete hadron need not exist while the field
acts. We use a simple Gaussian source model without assuming
that the constituents of $\X$ are $D$ and $\bar D^*$ mesons.
We retain the horizon distribution of Ref.~\cite{Yang} and
represent each horizon as the spatial range over which the field
acts coherently. A short phase pulse then changes the source-state
overlap, replacing the sharp survival condition used in the
bottomonium analysis. Averaging the pulse response over these
horizon lengths gives, to leading order in the phase variance,
a fractional loss of formation probability proportional to the
Gaussian overlap width. Extending this dependence to finite loss
is the attenuation assumption used below.

LHCb prompt data show that $\X/\Ps$ and $\Ps/\Jp$ both
increase with transverse momentum ($\pt$) over matched intervals
~\cite{Xproduction,PsiProduction,JpsiProduction,JpsiErratum}.
Within a common response model, their logarithmic changes measure
two adjacent differences in overlap width. Dividing the changes
removes the common field coefficient.
Potential-model separations for $\Jp$ and $\Ps$ then provide
reference scales for the correlations involved in $\X$ formation.
This comparison constrains an overlap width and the source needed
to resolve the states; it does not directly measure the full
$\X$ radius or its constituent composition.

We use published prompt yields, separated experimentally from
the contribution of displaced $b$-hadron decays
~\cite{CMSproduction,ATLASproduction,Xproduction}.
Since prompt $\Jp$ production includes feed-down from heavier
charmonia, we supplement the direct-$\Jp$ analysis with a
parent-mixture sensitivity check.
Earlier activity-dependent production analyses also show why
structural conclusions depend on the assumed interaction with the
environment~\cite{Xactivity,Compact,Molecule}.

We first describe the source-state projection and derive its
early-field response, including the assumption used to extend it
to finite attenuation. Charmonium activity data then check the
assumed $\pt$ dependence. We next use matched
three-state production ratios to constrain the source width and
the final $\X$ envelope, and discuss the production interpretation
and feed-down effects. Finally, we give a three-state activity
relation that tests the spatial response without assuming
a momentum dependence. Appendix~\ref{sec:responsechecks} derives
the nonlinear-response criterion and checks finite phase variance
and the effect of a nodal reference.

\FloatBarrier
\section{Early-field response and the causal horizon}
\label{sec:horizon}
\subsection{The source, the final state and their overlap}
\label{sec:source}
The three final species are $\Jp$, $\Ps$ and $\X$.
We use the labels $J$, $\psi$ and $3872$ for them throughout,
so the index $i\in\{J,\psi,3872\}$ identifies the final state.
Both the source and final envelope are functions of the same
internal relative-position vector $\mathbf x$.
For charmonium, $\mathbf x$ joins the heavy quark and antiquark;
for $\X$, it is an effective internal coordinate without
assigning constituents.

\Needspace{4\baselineskip}
The source has root-mean-square (RMS) separation $w$, and the
final-state envelope has RMS separation $r_i$.
All lengths refer to an effective precursor rest frame, in which
the source and field correlations are approximated as isotropic.

The source amplitude $S(\mathbf x)$ and final envelope $\psi_i(\mathbf x)$ are proportional to
\begin{equation*}
 \exp\!\left[-\frac{3\mathbf x^2}{4w^2}\right]
 \quad\text{and}\quad
 \exp\!\left[-\frac{3\mathbf x^2}{4r_i^2}\right],
\end{equation*}
respectively. Their squared amplitudes have RMS separations $w$
and $r_i$. The projection amplitude contains their product
$\psi_i^*S$. Its exponent is
$-3\mathbf x^2(w^{-2}+r_i^{-2})/4$, so the product is another
Gaussian with $s_i^{-2}=w^{-2}+r_i^{-2}$. Its overlap width is
\begin{equation}
 s_i=\frac{r_iw}{\sqrt{r_i^2+w^2}}.
 \label{eq:overlap}
\end{equation}
Here $w$ describes the source, $r_i$ the final envelope, and
$s_i$ the spatial extent of their overlap on which the field acts.
In particular, $s_i$ is a Gaussian width parameter, not the RMS
separation of the complete hadron.
For a broad source $s_i\simeq r_i$, while for a narrow source
$s_i\simeq w$ and the differences among the states become small.

For normalized source and final envelopes, their squared overlap
in the absence of the field is
\begin{equation*}
 \left|\int d^3x\,\psi_i^*(\mathbf x)\,S(\mathbf x)\right|^2
 =\left(\frac{2wr_i}{w^2+r_i^2}\right)^3.
\end{equation*}
This spatial factor is largest when the source and final envelope
have the same width. An envelope much broader than a fixed source
has a small projection, which can reduce the formation efficiency
even before the early field acts.
The cubic dependence follows from the Gaussian approximation;
a molecular wave function need not have the same dependence.
The field factor introduced below multiplies this no-field
projection, while an absolute yield also depends on the initial
production rate.

The Gaussian envelopes are nodeless. They omit the radial node
of the excited $\Ps$ state.
A radial node is a nonzero radius at which the radial wave function
vanishes and changes sign; the $1S$ $\Jp$ has no such node.
Spin and angular factors are omitted for all three species.
These envelopes do not form an orthogonal set of hadronic states,
so their squared overlaps cannot be summed as exclusive production
probabilities. Hard production and spin and orbital factors remain
in the production baseline.
The radial sign change allows phase fluctuations to alter both
destructive and constructive interference. The attenuation law inferred from
positive Gaussian envelopes need not hold for a full nodal
wave function. Appendix~\ref{sec:nodalcheck} gives an explicit
check with a radial node at the same reference RMS separations; it shows
that the RMS alone does not determine the phase response.

\subsection{A short phase pulse and spatial coherence}
We use a minimal spatial model: a prescribed Gaussian source
is exposed to a short phase pulse and then projected onto each
final-state envelope. During the pulse, internal motion is
neglected and the amplitude at $\mathbf x$ acquires the phase
$\ee^{-i\phi(\mathbf x)}$.
This sudden, phase-only approximation isolates the effect of
spatial interference. Spatial noise also enters open-quantum-system
and stochastic-potential treatments of quarkonium
~\cite{OQS,Stochastic,SpatialNoise}; the present model does not
calculate hard production or color and spin evolution.

The final-state probability contains interference between two
positions $\mathbf x$ and $\mathbf x'$ in the source-state overlap.
A common phase leaves that probability unchanged, while different
phases reduce the constructive interference of the positive
Gaussian envelopes. Let $F_i$ be the ratio of the formation
probability after the pulse to that for the same source without
the pulse.
Thus $F_i=0.7$ means a 30\% reduction in the probability of
forming species $i$.
This calculation isolates the pulse-induced change in the
spatial projection. It does not calculate subsequent binding,
rescattering or decay. A complete resonance need not have existed
during the pulse.

We take $\phi$ to be a zero-mean Gaussian random field.
Its dimensionless phase variance is
$\sigma_\phi^2\equiv\langle\phi(\mathbf x)^2\rangle$,
and its covariance is
\begin{equation}
 \langle\phi(\mathbf x)\phi(\mathbf x')\rangle
 =\sigma_\phi^2\exp\!\left[-\frac{|\mathbf x-\mathbf x'|^2}{2r_H^2}\right].
 \label{eq:covariance}
\end{equation}
Here $r_H$ is the \emph{causal horizon}~\cite{Yang}, represented
as the spatial correlation range of the field. It describes field
coherence and is distinct from the source width $w$.
The Gaussian covariance represents a smooth correlation range,
rather than a sharp spatial boundary. At separation
$d=|\mathbf x-\mathbf x'|$, averaging the relative phase gives
$\exp[-\sigma_\phi^2(1-\ee^{-d^2/(2r_H^2)})]$.

The product of source and final-state amplitudes is proportional
to $\exp[-3\mathbf x^2/(4s_i^2)]$.
For the phase average, the normalized weight is this amplitude
product, without an additional square. Its one-dimensional
Gaussian variance is $2s_i^2/3$ for each Cartesian component.
The probability contains two such weights, one at $\mathbf x$
and one at $\mathbf x'$, so their difference has variance
$4s_i^2/3$ in each coordinate. Its normalized three-dimensional
separation weight is consequently
\begin{equation*}
 \left(\frac{3}{8\pi s_i^2}\right)^{3/2}
 \exp\!\left[-\frac{3d^2}{8s_i^2}\right].
\end{equation*}
This is a density in $d^3(\mathbf x-\mathbf x')$; a radial
integral also includes the factor $4\pi d^2$.

\subsection{The horizon ensemble and the attenuation law}
We retain the exponential horizon distribution with a single
scale used in Ref.~\cite{Yang}, but replace its sharp survival
condition by the phase average of a source-state overlap.
The probability density of the causal horizon is
\begin{equation*}
 p(r_H)=\frac{\ee^{-r_H/\langle r_H\rangle}}{\langle r_H\rangle},
 \qquad r_H\geq0,
\end{equation*}
where $\langle r_H\rangle$ is the mean causal horizon.
For a weak pulse, the fractional loss satisfies
$-\ln F_i\simeq1-F_i$ to first order in $\sigma_\phi^2$.
The Gaussian spatial average of
$\exp[-d^2/(2r_H^2)]$ is
$(1+4s_i^2/(3r_H^2))^{-3/2}$.
The power $-3/2$ comes from the three Cartesian Gaussian integrals.
Averaging the remaining loss over the horizon distribution gives
\begin{align}
 -\ln F_i\simeq{}&\sigma_\phi^2\int_0^\infty
 \frac{dr_H}{\langle r_H\rangle}\ee^{-r_H/\langle r_H\rangle}
 \nonumber\\
 &\times\left[1-\left(1+\frac{4s_i^2}{3r_H^2}\right)^{-3/2}\right].
 \label{eq:horizonintegral}
\end{align}
For $s_i\ll\langle r_H\rangle$, introduce the dimensionless
integration variable $u=\sqrt{3}r_H/(2s_i)$, so that
$r_H=2s_i u/\sqrt{3}$. The remaining integral is
\begin{equation*}
 \int_0^\infty\left[1-\frac{u^3}{(1+u^2)^{3/2}}\right]du=2,
\end{equation*}
which gives the leading response
\begin{equation}
 -\ln F_i\simeq\frac{4\sigma_\phi^2}{\sqrt{3}\langle r_H\rangle}\,s_i.
 \label{eq:horizonresponse}
\end{equation}

The linear dependence has a physical origin. The ensemble includes
short coherent ranges that can resolve the overlap; their fraction
is proportional to $s_i/\langle r_H\rangle$.
At one fixed horizon much larger than the overlap, the leading
loss would instead be $2\sigma_\phi^2 s_i^2/r_H^2$.
The horizon distribution thus determines the spatial dependence
of the attenuation. Its exponential form is a model assumption,
rather than a consequence of causality alone.

Equation~\eqref{eq:horizonresponse} is valid to leading order
for small phase variance and overlap widths much smaller than
the mean horizon.
The formation-length comparison below does not require a numerical
value of $\langle r_H\rangle$. We adopt the leading length dependence
as an effective law at finite attenuation and absorb the unknown
normalization into a common coefficient:
\begin{equation}
 -\ln F_i=A\,s_i,
 \label{eq:linear}
\end{equation}
where the common positive coefficient $A$ has dimensions of inverse length.
In the leading limit, $A=4\sigma_\phi^2/(\sqrt{3}\langle r_H\rangle)$.
Extending the same spatial dependence to finite loss is an
additional phenomenological assumption. Saturation, finite variance
or a different distribution of short coherent ranges may change it.
In the three-state comparison, the ratio of logarithmic production
changes in Eq.~\eqref{eq:ruler} eliminates $A$.
Only the ratio of differences in overlap width remains, retaining
the spatial dependence supplied by the horizon ensemble.
An exact finite-variance pulse check is given in
Appendix~\ref{sec:finitecheck}. It does not make the finite-loss
law in Eq.~\eqref{eq:linear} a consequence of the weak-pulse
derivation.

\FloatBarrier
\section{Momentum dependence of charmonium ratios}
\label{sec:tests}
\subsection{Production ratios and proper-time exposure}
The production yield depends both on the initial production rate
and on the source-state overlap in the absence of the field.
To study changes in the field response, we compare the production
ratio of two species in a selected sample with that in a reference sample:
\begin{equation}
 C_{i/j}=\frac{[i/j]_{\rm selected}}{[i/j]_{\rm reference}}.
 \label{eq:contrast}
\end{equation}
Here $[i/j]$ is the prompt production ratio of species $i$ and $j$.
For example, $C_{i/j}=1.2$ means a 20\% increase relative to
the reference. Some published ratios contain decay branching
fractions. Because the same branching factors multiply both
samples, they cancel in the normalized change in Eq.~\eqref{eq:contrast}.

We first examine the momentum response using $\Jp$ and $\Ps$,
before drawing conclusions about the formation geometry of $\X$.
Assume a fixed field duration in a local longitudinal frame.
The proper time spent in that field decreases as the precursor
transverse energy increases.
If independent weak phase increments accumulate over this time,
$\sigma_\phi^2$ is proportional to the proper duration.
We use a common precursor mass $m_0=3.10$ GeV, set by $\Jp$.
At zero longitudinal momentum in this frame, its Lorentz factor
is $\sqrt{m_0^2+\pt^2}/m_0$, so the proper duration is
proportional to $m_0/\sqrt{m_0^2+\pt^2}$.
At fixed activity, the horizon distribution is assumed to be
independent of momentum in the spatial description above;
additional boost-dependent anisotropy of that description is
neglected. The exposure relative to a reference momentum $\pt^0$ is then
\begin{equation}
 g(\pt)=\frac{\sqrt{m_0^2+(\pt^0)^2}}
                    {\sqrt{m_0^2+\pt^2}}.
 \label{eq:exposure}
\end{equation}
It multiplies $A$ in Eq.~\eqref{eq:linear}.
This proper-time assumption supplies the momentum dependence
that is absent from the spatial pulse calculation.
The measured hadron momentum approximates the precursor momentum.
The same mass is used for both charmonium species and all
rapidity windows.

\subsection{Charmonium ratios versus charged-track multiplicity}
LHCb measured prompt $\Ps/\Jp$ versus the number of
primary-vertex tracks~\cite{CharmActivity}.
We compare the 20--45, 45--70, 70--95 and 95--200 track classes
with the 4--20 track reference at the same momentum and rapidity.
In this subsection, $C_{\psi/J}$ in Eq.~\eqref{eq:contrast}
therefore means an activity change, not a momentum change.

For each class, the measured ratio at $0.3<\pt<2$ GeV/$c$
and $2.8<y<3.5$ provides the low-momentum input.
For example, the published ratios in 20--45 and 4--20 tracks
are 10.92\% and 15.18\%, giving a normalized value
$10.92/15.18\simeq0.72$.
If activity changes the field coefficient while the source and
relative production baseline remain fixed, the momentum dependence
of this normalized activity ratio is
\begin{equation}
 C_{\psi/J}(N,\pt)
       =[C_{\psi/J}(N,\pt^0)]^{g(\pt)},
 \label{eq:activity}
\end{equation}
where $N$ labels the track class and
$\pt^0=1.15$ GeV/$c$ is the reference-bin midpoint.
The measured low-momentum ratios set the track-class dependence;
the model does not convert track counts into horizon lengths.

\begin{figure}[H]
\includegraphics[width=\textwidth]{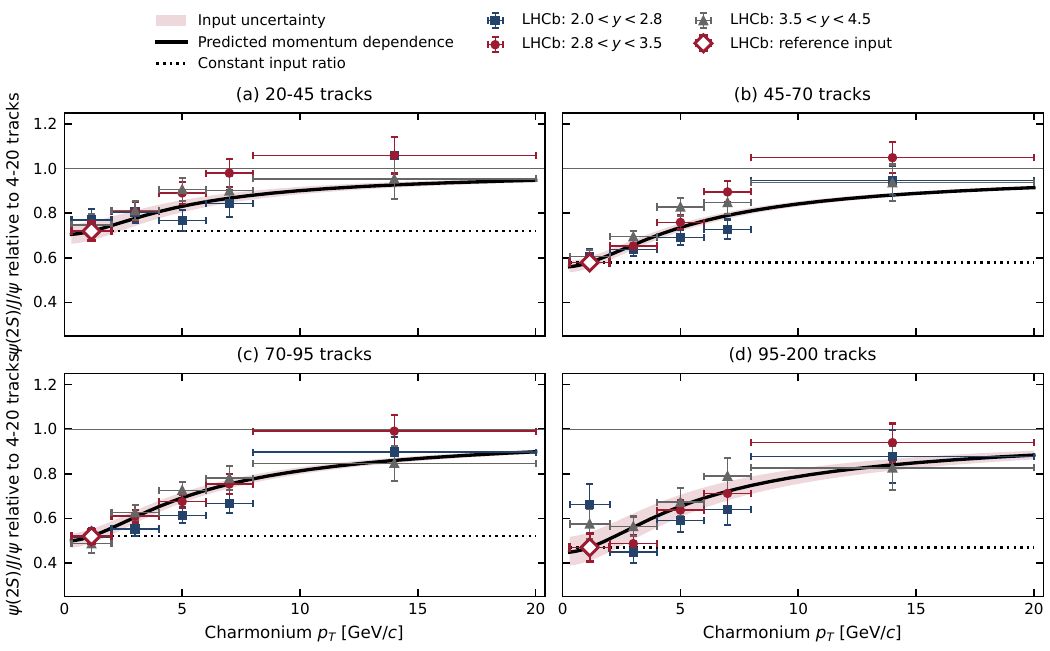}
\caption{Checking the momentum dependence with LHCb prompt
$\Ps/\Jp$ activity data~\cite{CharmActivity}.
Each panel compares the indicated track class with 4--20 tracks
at the same kinematics. The open diamond marks the
low-momentum input in $2.8<y<3.5$.
Solid curves use Eq.~\eqref{eq:activity}; dotted curves keep
the input ratio constant with momentum.
Bands propagate the uncertainty of this input.
Other points test the momentum dependence and its applicability
to the other rapidity windows. Horizontal bars show momentum
intervals; vertical bars combine published uncertainty components
in quadrature for display.}
\label{fig:activity}
\end{figure}

Figure~\ref{fig:activity} shows that this rule captures the
principal recovery toward unity as momentum increases across
the activity classes and rapidity windows.
The dotted curves retain the low-$\pt$ suppression at every
$\pt$, whereas the data approach unity as $\pt$ increases.
No additional momentum or rapidity parameters are used.
The approximation does not calculate rapidity-dependent field
evolution or changes in the prompt-parent composition.
We evaluate Eq.~\eqref{eq:activity} at momentum-bin midpoints.
For the wide $8<\pt<20$ GeV/$c$ interval this means
$\pt=14$ GeV/$c$, which approximates a spectrum-weighted bin average.
The normalized points share reference bins, so their errors are
correlated. This comparison tests the main trend and is not used
for a precision goodness-of-fit assessment.

The field model does not calculate the low-momentum ratios that
set the size of the multiplicity effect. The remaining points
test their momentum dependence and its applicability across
rapidity windows. Since each input ratio absorbs the difference
between the state responses, this check does not determine the
spatial attenuation law.

\FloatBarrier
\section{Inferring X(3872) formation scales}
\label{sec:data}
\subsection{Removing the common field coefficient}
We now compare the spatial responses of the three states to
constrain $\X$ formation. This step does not require the
momentum rule in Eq.~\eqref{eq:exposure}: dividing the logarithmic
changes of adjacent production ratios cancels the common field
coefficient for any momentum dependence.

We assume a common source width $w$ for the three species,
with both $w$ and their relative production baselines fixed
between the compared samples. The baseline includes hard
production and the source-state overlap without the field.
The species may have different baselines, but their ratios
must remain unchanged. We also assume a common effective $A$
for the three species in each inclusive bin; averaging over
different event populations does not guarantee a common response.
Changes in hard production can contribute to the momentum
dependence, so the ratio data alone do not establish the
fixed-baseline assumption.
The common coefficient also assumes a common field coupling to
the effective precursor coordinate. Since the complete hadron
need not exist during the pulse, this is a common-precursor
response assumption, rather than a calculation of each final
hadron's QCD coupling.
Under these conditions, Eq.~\eqref{eq:linear} gives
\begin{equation*}
 \ln C_{i/j}=(A_{\rm reference}-A_{\rm selected})(s_i-s_j).
\end{equation*}
Dividing the logarithmic changes of the two adjacent ratios
removes the common coefficient:
\begin{equation}
 \lambda\equiv\frac{\ln C_{3872/\psi}}{\ln C_{\psi/J}}
 =\frac{s_{3872}-s_\psi}{s_\psi-s_J}.
 \label{eq:ruler}
\end{equation}
The mean horizon and field strength cancel through the common
coefficient, but the spatial dependence remains in the differences
of $s_i$. A quadratic spatial law, for example, would give
differences of $s_i^2$. The three-state comparison therefore uses
the spatial law supplied by the horizon ensemble without requiring
the numerical horizon scale.
If the $\X$ attenuation coefficient instead contains an unknown
relative strength $\kappa$, the numerator on the right of
Eq.~\eqref{eq:ruler} becomes $\kappa s_{3872}-s_\psi$.
The ratios would then determine $\kappa s_{3872}$ rather than
$s_{3872}$ alone. We do not fit this extra factor: the length
conversion below assumes $\kappa=1$ and therefore gives
effective formation scales under common field coupling.

\subsection{Matched LHCb data and a common-slope fit}
We combine the prompt $\Ps$ cross sections in Table 2 of
Ref.~\cite{PsiProduction}, the corrected prompt $\Jp$ cross
sections~\cite{JpsiProduction,JpsiErratum}, and the prompt
$\X/\Ps$ ratios in Fig. 4 of Ref.~\cite{Xproduction}.
All three measurements were made in proton-proton collisions
at $\sqrt{s}=13$ TeV. Their common collision energy allows
a direct comparison of the momentum dependences.
The common $\pt$ intervals are 4--6, 6--7, 7--8,
8--9, 9--10, 10--12 and 12--14 GeV/$c$, in the rapidity windows
$2<y<3$, $3<y<3.5$ and $3.5<y<4.5$.
The published $\Jp$ and $\Ps$ cross sections are integrated
over complete bins to match these intervals, without interpolation.
The $\X/\Ps$ values and total uncertainties are digitized
from Fig. 4 of Ref.~\cite{Xproduction}.
The reconstruction details are documented with the analysis files.

We determine the spatial slope from all 21 matched points, with
one common $\lambda$ for the three rapidity windows.
Equation~\eqref{eq:ruler} is equivalent to
\begin{equation*}
 \ln[\X/\Ps]=a_y+\lambda\ln[\Ps/\Jp],
\end{equation*}
where the constant $a_y$ sets the overall normalization in each
rapidity window and absorbs fixed branching fractions.
The measured $\Ps/\Jp$ ratio carries the $\pt$ variation
on the right-hand side, while $\lambda$ determines the
corresponding variation of $\X/\Ps$.
The fit has one common slope and three constant intercepts.

We include errors in both measured ratios. At each trial slope,
the three normalizations and the true horizontal coordinates are
profiled, meaning that they are chosen to minimize $\chi^2$.
The main fit treats the quoted total errors of the original,
unnormalized points as independent. In logarithmic
coordinates, a residual then has variance
$\sigma^2_{\ln(\X/\Ps)}+\lambda^2\sigma^2_{\ln(\Ps/\Jp)}$.
Fitting before normalization retains the correlations that would
arise from dividing several points by a common reference bin.
It gives
\begin{equation*}
 \lambda=0.87^{+0.17}_{-0.16},\qquad \chi^2=8.9\quad
 \text{for 17 degrees of freedom}.
\end{equation*}
The errors correspond to a unit increase in the profiled $\chi^2$
under this stated error model. As a sensitivity check, treating
the published correlated charmonium systematic components as
coherent shifts changes the central slope to 0.86.
The full experimental covariance of these measurements is not
available. The quoted fit uncertainty
therefore does not include every experimental correlation.

As an illustration, the change from 4--6 to 12--14 GeV/$c$
in $2<y<3$
gives $C_{3872/\psi}=0.113/0.072\simeq1.57$ and
$C_{\psi/J}=0.3118/0.1928\simeq1.62$.
The ratio of these logarithmic changes is 0.94, consistent with
the common fit; it is not an additional fit input.

For display, we normalize to the 4--6 GeV/$c$ interval in each
rapidity window. The fitted relation then reads
\begin{equation}
 C_{3872/\psi}(\pt,y)=[C_{\psi/J}(\pt,y)]^\lambda.
 \label{eq:transfer}
\end{equation}

\begin{figure}[!htbp]
\includegraphics[width=\textwidth]{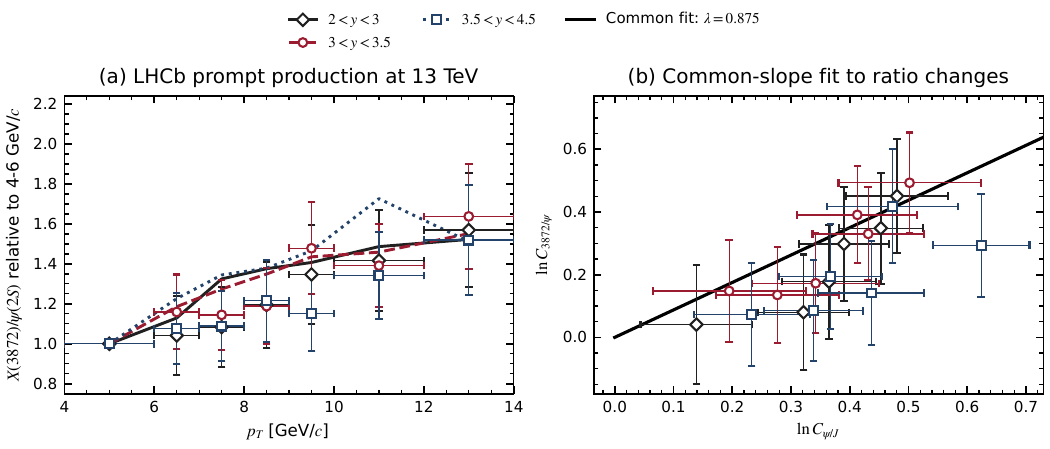}
\caption{Matched LHCb prompt production ratios at 13 TeV
~\cite{Xproduction,PsiProduction,JpsiProduction,JpsiErratum}.
(a) $\X/\Ps$ normalized to 4--6 GeV/$c$ in each rapidity window.
The black solid, red dashed and blue dotted curves correspond
to $2<y<3$, $3<y<3.5$ and $3.5<y<4.5$, respectively.
They apply Eq.~\eqref{eq:transfer} to the measured $\Ps/\Jp$
changes at the common fitted slope.
They need not be straight as functions of $\pt$; the straight
relation is between logarithmic ratio changes in panel (b).
(b) Logarithmic changes of the two ratios and the common fitted
relation. All matched points enter the fit; the lines are not
independent predictions for these data.
Horizontal bars in (a) show momentum intervals. For display,
the reported uncertainty components are combined in quadrature.
The fit includes errors in both ratios and retains the effect of
shared normalization bins.}
\label{fig:ruler}
\end{figure}

\Needspace{4\baselineskip}
Figure~\ref{fig:ruler} shows that one slope describes the main
rise across all three rapidity windows within the adopted errors.
Some intermediate points lie below the relation. Allowing separate
slopes for the three rapidity windows reduces $\chi^2$ by only 1.6
for two additional parameters.
The data therefore do not require a rapidity-dependent spatial slope.
We use the central common fit to infer formation scales below.
Assigning confidence levels to these scales would also require
the experimental correlations and model uncertainties.

\FloatBarrier
\section{Formation scales of X(3872)}
\label{sec:interpretation}
\subsection{Constraints on the source and the final envelope}
We calibrate the internal coordinate with the potential-model RMS
heavy-quark separations $r_J=0.50$ fm and $r_\psi=0.90$ fm
from Ref.~\cite{Satz}.
These are quark-antiquark separations, not constituent radii
about the center of mass. The inferred $r_{3872}$ uses the same
Gaussian convention. We first use a direct-$\Jp$ reference;
parent decays in the prompt sample are treated in
Sec.~\ref{sec:parents}.

The common slope supplies one spatial constraint on the two
unknown widths $w$ and $r_{3872}$. We therefore infer an allowed
curve and limiting scales, rather than two independently measured
lengths.

The positive central slope requires $s_{3872}>s_\psi>s_J$.
At fixed $w$, $s(r,w)$ is increasing and concave: each additional
unit of final separation adds progressively less overlap width.
Equation~\eqref{eq:ruler} consequently implies
\begin{equation}
 r_{3872}\geq r_\psi+\lambda(r_\psi-r_J)\simeq1.25\ {\rm fm}.
 \label{eq:profilebound}
\end{equation}
This minimum is approached as the source becomes arbitrarily
broad. A finite source requires a larger final envelope.

There is a separate source constraint. Since $s_{3872}<w$
for any finite final envelope,
\begin{equation}
 \lambda<\frac{w-s_\psi}{s_\psi-s_J}.
 \label{eq:sourcebound}
\end{equation}
For $w$ much smaller than either charmonium separation,
Eq.~\eqref{eq:overlap} gives
$s_i=w-w^3/(2r_i^2)+\mathcal O(w^5/r_i^4)$.
Substituting this expansion into Eq.~\eqref{eq:sourcebound}
shows that its right-hand side approaches
\begin{equation*}
 \frac{r_J^2}{r_\psi^2-r_J^2}
 =\frac{0.50^2}{0.90^2-0.50^2}\simeq0.45.
\end{equation*}
Thus a very narrow source cannot give a slope above about 0.45,
even for an arbitrarily extended final $\X$ envelope.
The central fitted slope, 0.87, exceeds this maximum.
Solving Eq.~\eqref{eq:sourcebound} therefore requires $w>0.600$ fm.
For a narrower common source, the overlap widths of the three
states become too similar to reproduce the fitted slope, even
if the final $\X$ envelope is made arbitrarily large.

\begin{figure}[!htbp]
\centering
\includegraphics[width=0.62\textwidth]{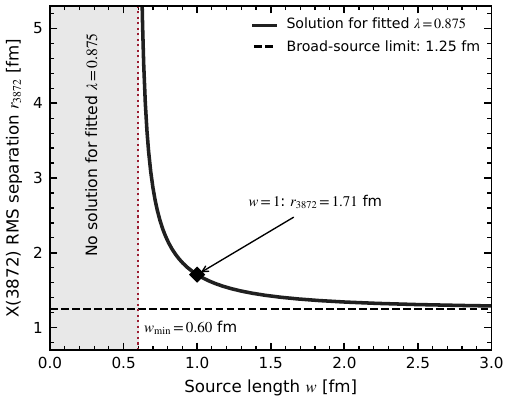}
\caption{Source dependence of the $\X$ RMS separation in the
common Gaussian model, using the central fitted $\lambda=0.87$ and
a direct-$\Jp$ reference. The curve solves Eq.~\eqref{eq:ruler}.
The shaded strip has no solution for this central slope.
The dashed line is the broad-source minimum, 1.25 fm;
the diamond marks the illustrative choice $w=1$ fm.
These central model solutions have no assigned confidence level.}
\label{fig:boundary}
\end{figure}

\subsection{Dependence on the assumed source width}
No source width is fixed to obtain the two minimum scales.
Figure~\ref{fig:boundary} and Table~\ref{tab:source} show how
the solution changes as $w$ varies.
In particular, $w=1$ fm gives $r_{3872}\simeq1.71$ fm,
but this is only one point on the curve. Near the source threshold
the inferred $r_{3872}$ becomes very large; for broad sources
it approaches 1.25 fm.
The inferred $r_{3872}$ depends on the source width.
Within this Gaussian model, the production-ratio data place
no finite upper bound on the final-envelope separation.

Even when the final envelope becomes very broad, its overlap
with a finite source saturates and remains restricted. From Eq.~\eqref{eq:ruler},
$s_{3872}=(1+\lambda)s_\psi-\lambda s_J$.
As $w$ increases, this combination increases monotonically.
At the source threshold it approaches 0.600 fm, and for a broad
source it approaches 1.250 fm. Thus the central direct-reference
solution gives an overlap-width interval of approximately
0.60--1.25 fm, without choosing $w=1$ fm.
This is a constraint on the Gaussian overlap sampled by the field,
not an interval for the RMS separation of the complete $\X$.

\begin{table}[!htbp]
\centering
\caption{Source-width dependence at the central fitted slope, using a direct-$\Jp$
reference. All lengths are in fm. The last row is the broad-source limit;
none of the finite source widths is fitted or selected by the data.}
\label{tab:source}
\begin{tabular}{ccc}
\toprule
Source $w$ (fm) & Final separation $r_{3872}$ (fm) & Overlap width $s_{3872}$ (fm)\\
\midrule
0.70 & 2.87 & 0.68\\
0.80 & 2.16 & 0.75\\
1.00 & 1.71 & 0.86\\
2.00 & 1.34 & 1.11\\
$\infty$ & 1.25 & 1.25\\
\bottomrule
\end{tabular}
\end{table}

These source-width variations apply within the specified Gaussian
model. Uncertainty in the spatial response law, reference separations
and production baselines also affects the scales.
The source lower bound depends on the fitted slope:
it disappears if $\lambda$ falls below the narrow-source
limit of about 0.45.
We quote central model scales, without assigning a confidence
level to the length bounds.

The lower bounds have a useful extension beyond a strictly linear
response. They remain valid when the difference in logarithmic
attenuation between the two samples is increasing and concave
as a function of the Gaussian overlap width
(Appendix~\ref{sec:concavecheck}). For this class, the curve in
Fig.~\ref{fig:boundary} becomes a lower boundary, rather than an
equality solution. The finite upper bound on $s_{3872}$ need not
survive. The exact pulse does not guarantee this shape condition
at all field strengths, and this extension still uses positive
Gaussian envelopes. It is not a wave-function-independent radius
bound.

\subsection{What can be inferred about X(3872)?}
The fitted $\lambda\simeq0.87$ means that the additional
overlap width from $\Ps$ to $\X$ is almost as large as the
width difference from $\Jp$ to $\Ps$.
Within the shared-source response model, formation of $\X$
therefore involves spatial correlations that the early field can
distinguish from those of $\Ps$.
The lower bound on $w$ gives the minimum spatial extent of the
source consistent with this relative response. Together with the
bound on the final envelope, it constrains the initial correlations
that can lead to the measured formation trend. This joint source
and final-state constraint is the main physical result within
the positive-envelope response model.

Under a $D\bar D^*$ molecular interpretation, a localized
production source must project onto the relative meson wave
function. If that wave function extends well beyond the prepared
region, its normalization can leave only a small amplitude where
the source is appreciable. The spatial overlap discussed in
Sec.~\ref{sec:source} can then suppress the formation efficiency
per prepared pair. A small prompt yield and an extended bound
state can therefore belong to the same formation picture.
This picture applies both to an extended molecular wave function
and to an extended state with a compact
component~\cite{Factorization,ProductionScales,CoupledChannel}.

Individual meson sizes do not set a sharp formation cutoff.
For example, a lattice-QCD calculation gives charged-meson RMS
charge radii of about 0.37 fm for $D^+$ and 0.43 fm for
$D^{*+}$~\cite{DMesonRadius}. These describe the internal charge
distributions. Their sum is neither the width of the produced
$D\bar D^*$ source nor a maximum separation for forming a
molecule. A quantitative molecular calculation requires the
relative wave function and its coupling to the early field,
including the relevant short-distance dynamics.

An absolute yield also depends on the production rate of suitable
precursors and on their spin and momentum distributions.
Here we fit normalized changes of production ratios. Under the fixed-source
assumption, a constant no-field overlap is part of the production
baseline and cancels from these changes. The fit therefore
constrains how the early field modifies formation; it does not
determine how much of the small absolute yield is caused by
geometric overlap suppression. Indeed, the allowed Gaussian
curve crosses $w=r_{3872}$, where the no-field spatial overlap
is unity. The scale constraints permit a small baseline overlap
but do not require it.

Using $\Jp$ as a reference does not remove the species-dependent
production baseline. If $[\X/\Jp]_0$ denotes the ratio without
the early field, the model gives
\begin{equation*}
 [\X/\Jp]=[\X/\Jp]_0
 \exp[-A(s_{3872}-s_J)].
\end{equation*}
The unknown baseline ratio contains hard-production, spin and
overlap factors. A measured ratio in one reference bin can
normalize predictions for relative changes between bins, but
does not determine $[\X/\Jp]_0$.

The resulting formation constraint complements line-shape and
weak-binding analyses, which use threshold amplitudes, scattering
parameters and pole information~\cite{Lineshape,RecentPole,WeakBinding}.
The Gaussian overlap width cannot be identified with the full
molecular separation or converted directly into a compositeness
probability. It also does not distinguish a bound pole from
a virtual pole.

A compact final envelope requires a momentum-dependent production
baseline. For $r_{3872}\leq r_\psi$, the same fixed source
and a field coefficient that decreases as momentum rises cannot
produce an increasing $\X/\Ps$ ratio.
The relative $\X/\Ps$ production baseline must then increase
by at least the observed factor $0.113/0.072\simeq1.57$, or 57\%,
from 4--6 to 12--14 GeV/$c$ in $2<y<3$
~\cite{CoreProduction}.

\subsection{Magnitude of the early-field response}
\label{sec:fieldstrength}
The magnitude of the measured ratio changes also constrains
the formation response. Using the direct-$\Jp$ approximation,
we return to the comparison from 4--6 GeV/$c$ (``low'') to 12--14 GeV/$c$ (``high'')
in $2<y<3$. With the same fixed source and relative production
baselines, the model assigns this change to a decrease in $A$.

\Needspace{7\baselineskip}
Using $s_{3872}=s_\psi+\lambda(s_\psi-s_J)$ gives
\begin{align*}
 -\ln\frac{F_{3872}^{\rm low}}{F_{3872}^{\rm high}}
 &=\ln C_{\psi/J}
   \left[\frac{s_\psi}{s_\psi-s_J}+\lambda\right]\\
 &\geq\ln C_{\psi/J}
   \left[\frac{r_\psi}{r_\psi-r_J}+\lambda\right].
\end{align*}
The smallest value occurs for a broad source. The measured
$C_{\psi/J}\simeq1.62$ and fitted $\lambda\simeq0.87$ give
a central minimum logarithmic change of 1.50, or
$F_{3872}^{\rm low}/F_{3872}^{\rm high}\lesssim0.22$.
Thus, at the central fit, the attenuation factor $F_{3872}$
must increase by at least a factor of about 4.5 between these
two momentum intervals. Here $F_{3872}$ is the formation probability
relative to its no-field value. The increase follows from the
assumed response law; it is neither a measured cross-section
ratio nor an independent determination of an absolute probability.
Because the attenuation law has $A\geq0$, its high-$\pt$ factor
satisfies $F_{3872}^{\rm high}\leq1$. The same result therefore gives
\begin{equation*}
 F_{3872}^{\rm low}\lesssim0.22.
\end{equation*}
At the central fit, at most about 22\% of the no-field formation
probability remains in the 4--6 GeV/$c$ bin under these assumptions.
A measured yield in that bin would imply a no-field production
baseline at least about 4.5 times larger, rather than an upper
limit on the absolute $\X$ cross section.

A change of this size is outside the weak-pulse limit.
The numerical result therefore uses the effective finite-loss
law in Eq.~\eqref{eq:linear}, with the same source, baseline,
coupling and reference assumptions as the length constraints.
The lower change also follows under the nonlinear shape condition
of Appendix~\ref{sec:concavecheck} if the logarithmic attenuation
difference vanishes at zero width. An arbitrary finite pulse need
not satisfy this condition.

\subsection{Feed-down in the prompt reference}
\label{sec:parents}
Prompt $\Jp$ includes direct production and decays of $\chi_c$
and $\Ps$ produced at the collision. The early field acts on
the parent states before they decay.
To examine a finite change in attenuation, define $f_a^{\rm high}$
as the observed fraction of parent $a$ in the high-$\pt$ reference
bin, with $\sum_a f_a^{\rm high}=1$.
Let $\Delta A=A_{\rm low}-A_{\rm high}>0$.
At a fixed source width, each parent's low-to-high yield change
contains $\exp(-\Delta A s_a)$ and a common production factor.
That common factor cancels from the species ratio, giving
\begin{equation}
 C_{\psi/J}
 =\sum_a f_a^{\rm high}\exp[\Delta A(s_\psi-s_a)].
 \label{eq:parents}
\end{equation}
Here $a$ runs over direct $\Jp$, $\chi_c$ and $\Ps$.
The sum contains their individual overlap widths, and retains the
change of the parent mixture at finite attenuation.
It neglects momentum migration in the decays.

For illustration, we assign the historical
fractions $(0.60,0.30,0.10)$~\cite{Satz} to the high-$\pt$ bin,
with parent RMS separations $(0.50,0.72,0.90)$ fm.
These are not measured fractions of the matched 13 TeV sample.
Using $C_{\psi/J}=1.6168$ for the illustrated endpoint pair fixes
$\Delta A$ at each $w$ through Eq.~\eqref{eq:parents}.
Using the central slope in Eq.~\eqref{eq:transfer} then gives
\begin{equation*}
 s_{3872}=s_\psi+\frac{\lambda\ln C_{\psi/J}}{\Delta A}.
\end{equation*}
This is an endpoint sensitivity calculation using the existing
common slope, not a new fit to the 21 prompt points.
It does not require the absolute value of $A_{\rm high}$.

\begin{table}[H]
\caption{Central formation scales for the direct reference and
a finite parent-mixture example. All lengths are in fm.
The parent example uses fractions specified in the high-$\pt$
bin and the illustrated endpoint amplitude; it is not a measured
composition or a full feed-down correction.}
\label{tab:sensitivity}
\centering
\begin{tabular}{lccc}
\toprule
Reference & Minimum & Minimum & $r_{3872}$ at\\
treatment & $r_{3872}$ (fm) & $w$ (fm) & $w=1$ fm\\
\midrule
Direct $\Jp$ & 1.25 & 0.60 & 1.71\\
Finite parent example & 1.17 & 0.40 & 1.41\\
\bottomrule
\end{tabular}
\end{table}

The final-envelope scale remains extended in this example,
while the source lower scale is more sensitive to the reference
treatment. The illustrative fractions, their momentum dependence
and decay migration must be determined for a quantitative
correction. Different source widths or changing production
baselines could also modify the scales; neither is fitted here.
The Gaussian reference shape is a separate limitation:
Appendix~\ref{sec:nodalcheck} shows explicitly that a nodal
reference with the same RMS can have a different phase response.

\FloatBarrier
\section{A prediction for matched activity data}
\label{sec:future}
A matched activity measurement can test the spatial response
without using the momentum rule $g$.
The three prompt species must be measured at the same collision
energy and in the same momentum, rapidity and activity intervals.
Normalizing each species ratio to a reference track class as in
Eq.~\eqref{eq:contrast} gives the exact identity
\begin{equation*}
 C_{3872/J}=C_{3872/\psi}\,C_{\psi/J}.
\end{equation*}
This follows from $[\X/\Jp]=[\X/\Ps][\Ps/\Jp]$
in both the selected and reference samples.
If activity changes a common field coefficient while the source
and relative production baselines stay fixed, the same measured
spatial slope predicts
\begin{align}
 C_{3872/\psi}&=(C_{\psi/J})^{\lambda},\nonumber\\
 C_{3872/J}&=(C_{\psi/J})^{1+\lambda}.
 \label{eq:future}
\end{align}
For example, a measured $C_{\psi/J}=0.70$ and the central
$\lambda=0.87$ give $C_{3872/\psi}\simeq0.73$ and
$C_{3872/J}\simeq0.51$.
The exponent $1+\lambda$ therefore follows by multiplying
$C_{3872/\psi}=(C_{\psi/J})^\lambda$ by $C_{\psi/J}$;
it introduces no further dynamical assumption.
One measured $\Ps/\Jp$ change thus fixes both changes
involving $\X$, without another activity coefficient.
This equality prediction uses the linear effective law in
Eq.~\eqref{eq:linear}. The nonlinear concavity condition alone
preserves lower bounds, not a universal slope between environments.
The ordering of overlap widths predicts that both adjacent ratios
rise as exposure decreases with momentum and fall as the field
response strengthens with activity.

The activity comparison must treat feed-down consistently with
the momentum analysis. Parent-resolved or feed-down-corrected
$\Jp$ yields, cross-bin and cross-species covariance, and
nonprompt control samples would make the test more informative.
The existing 8 TeV $\X/\Ps$ activity result~\cite{Xactivity}
and 13 TeV $\Ps/\Jp$ result~\cite{CharmActivity}
are not a matched three-state measurement.

\FloatBarrier
\Needspace{8\baselineskip}
\section{Conclusions}
The matched LHCb prompt ratios at 13 TeV admit a common
logarithmic slope $\lambda\simeq0.87$ across three rapidity windows.
In a model with a common source, common field coupling and
fixed relative production baselines, this relates two adjacent
differences in the response to the early field.
For positive Gaussian envelopes and a direct-$\Jp$ reference,
the central fit requires a final $\X$ Gaussian RMS separation
of at least about 1.25 fm and a source RMS width above about
0.60 fm. The constraints concern the source and final envelope
together: increasing the final tail cannot compensate for an
arbitrarily narrow source within this response model.

The causal-horizon ensemble supplies the leading linear
dependence on overlap width. Extending it to finite attenuation
allows the three-state comparison to eliminate the common field
coefficient, without independently measuring the horizon scale.
The lower bounds also survive a class of nonlinear responses
with an increasing, concave logarithmic attenuation difference.
The exact pulse does not guarantee this condition globally,
and the finite overlap-width upper bound does not follow from
the nonlinear generalization.

The length conversion depends on the reference response.
A finite parent-mixture example gives a final-envelope lower
scale near 1.17 fm and a source threshold near 0.40 fm; matched
parent fractions and decay migration remain to be determined.
The horizon average probes separations between interfering
positions in the source-state overlap. The fixed nodal-reference
check shows that RMS separation alone does not determine this
interference-weighted response.
Thus the quoted lengths are effective scales of the stated
positive-envelope, common-coupling model, with no assigned
confidence level. They do not determine the molecular fraction,
binding energy or pole nature of $\X$.

For the central fit and direct reference, the same assumptions
give at least a factor of about 4.5 increase in the $\X$
formation factor from 4--6 to 12--14 GeV/$c$ in $2<y<3$,
and a low-$\pt$ factor at most about 0.22 relative to its
no-field value. Absolute production yields and the contribution
of baseline overlap suppression remain undetermined.

The charmonium activity comparison tests the assumed momentum
dependence after measured inputs fix its amplitude.
A matched three-state activity measurement would test the
spatial relation in Eq.~\eqref{eq:future} without another
activity-dependent coefficient. Parent-resolved references and
experimental covariance would make that test more decisive.
Extensions to $Y$ states require appropriate reference states
and matched production data~\cite{XYZreview}.

\FloatBarrier
\appendix
\setcounter{equation}{0}
\renewcommand{\theequation}{\thesection\arabic{equation}}
\renewcommand{\theHequation}{appendix.\arabic{equation}}
\section{Checks of the spatial response}
\label{sec:responsechecks}
\subsection{A sufficient condition for nonlinear attenuation}
\label{sec:concavecheck}
Retain a common source, fixed relative baselines, common field
coupling and positive Gaussian envelopes.
For a pair of momentum bins define the logarithmic attenuation
difference
\begin{equation}
 D(s)=\ln\frac{F_{\rm high}(s)}{F_{\rm low}(s)},\qquad
 L=D(s_\psi)-D(s_J)=\ln C_{\psi/J}>0.
 \label{eq:shapedifference}
\end{equation}
For that pair, $\lambda=[D(s_{3872})-D(s_\psi)]/L$.
If $D$ is increasing and concave on the sampled interval,
its successive secant slopes decrease. Hence
\begin{equation}
 \lambda\leq\frac{s_{3872}-s_\psi}{s_\psi-s_J}.
 \label{eq:shapebound}
\end{equation}
Since $s(r,w)$ is itself increasing and concave, the final-envelope
bound in Eq.~\eqref{eq:profilebound} follows. Since $s_{3872}<w$,
the source condition in Eq.~\eqref{eq:sourcebound} also follows.
For nonlinear $D$, the linear model gives a lower boundary
rather than an equality curve.

If concavity holds from zero width and $D(0)=0$, the secant
from zero to $s_J$ is at least that from $s_J$ to $s_\psi$.
Consequently
\begin{align}
 D(s_J)&\geq\frac{L s_J}{s_\psi-s_J},\nonumber\\
 D(s_{3872})&=D(s_J)+(1+\lambda)L\nonumber\\
 &\geq L\left[\frac{s_\psi}{s_\psi-s_J}+\lambda\right]
 \geq L\left[\frac{r_\psi}{r_\psi-r_J}+\lambda\right].
 \label{eq:shapeloss}
\end{align}
This preserves the lower field-factor change in
Sec.~\ref{sec:fieldstrength} under the stated shape conditions.
It does not preserve the finite upper bound on $s_{3872}$.
Applying the common fitted slope gives central conditional
scales; no new statistical fit or confidence assignment is made.

\subsection{Exact finite-variance pulse}
\label{sec:finitecheck}
The exact Gaussian phase average in Sec.~\ref{sec:horizon}
can be evaluated without expanding in $\sigma_\phi^2$.
With $t=\sqrt{2/3}\,s/\langle r_H\rangle$, expanding only the
inner exponential gives the convergent representation
\begin{align}
 F(t,\sigma_\phi^2)&=e^{-\sigma_\phi^2}\sum_{n=0}^\infty\frac{(\sigma_\phi^2)^n}{n!}f_n(t),
 \nonumber\\
 f_n(t)&=\int_0^\infty e^{-v}
 \left(\frac{v^2}{v^2+2nt^2}\right)^{3/2}dv,
 \qquad n\geq1,\quad f_0=1.
 \label{eq:exactpulse}
\end{align}
Here $v=r_H/\langle r_H\rangle$ is dimensionless.
An independent Laplace representation is
\begin{equation*}
 f_n(t)=\int_0^\infty e^{-tu}
 \frac{6n u^2}{(u^2+2n)^{5/2}}du.
\end{equation*}
Thus each $F$ is a Laplace transform of a positive normalized
measure, including the $n=0$ term. Its logarithmic loss is
concave in $t$, but the difference of two such losses need not be.
For example, $\sigma_{\phi,{\rm low}}^2=20$, $\sigma_{\phi,{\rm high}}^2=10$ and
$t=10$ give $D''(t)=+0.00910$ by both integral representations
and direct finite differences. This constructed counterexample
is outside the small-overlap limit; it disproves global concavity,
not the applicability of a particular finite-loss region.

The condition can also hold at finite loss. For example,
$(\sigma_{\phi,{\rm low}}^2,\sigma_{\phi,{\rm high}}^2)=(5,1)$ gives
$D(1)=1.78$ and $D''(1)=-0.646$.
Both examples use fixed variances to check the spatial response;
they are not fits to production data or determinations of a
momentum evolution. The sufficient shape condition can therefore
be useful beyond a weak pulse, but must be checked in the
range used for a physical application. We do not fit additional
phase-variance or horizon parameters to validate the length bounds.

\subsection{A fixed nodal-reference check}
\label{sec:nodalcheck}
To test the use of RMS separations alone, take a radial $1S$
reference and an excited-state radial wave function with one
radial node:
\begin{equation}
 \psi_J(\mathbf x)\propto e^{-a_J\mathbf x^2},\qquad
 \psi_\psi(\mathbf x)\propto(1-b\mathbf x^2)e^{-a\mathbf x^2}.
 \label{eq:nodalreference}
\end{equation}
The second wave function vanishes at
$|\mathbf x|=1/\sqrt b$ and has opposite signs on the two sides
of that radius.
The reference $r_J=0.50$ fm fixes $a_J=3.00$ fm$^{-2}$.
Orthogonality fixes $b=2(a+a_J)/3$, and imposing
$r_\psi=0.90$ fm on the normalized squared amplitude gives
$a=2.353$ fm$^{-2}$ and $b=3.569$ fm$^{-2}$.
These numbers are fixed by the two existing RMS inputs, with
no production or field parameter fitted. This is a shape
diagnostic, not a potential-model charmonium calculation.

\Needspace{5\baselineskip}
Normalize the amplitude product as
$W_\psi(\mathbf x)=\psi_\psi^*(\mathbf x)S(\mathbf x)/
\int d^3x\,\psi_\psi^*S$.
This interference weight can change sign.
For a pair separation $d$, the horizon integral satisfies
\begin{equation*}
 \int_0^\infty\left[1-e^{-d^2/(2r_H^2)}\right]dr_H
 =\sqrt{\frac{\pi}{2}}\,d.
\end{equation*}
Thus the leading loss probes a weighted separation between
interfering positions. Writing
$-\ln F_\psi\simeq4\sigma_\phi^2\ell_\psi/(\sqrt3\langle r_H\rangle)$
in the same weak-pulse, broad-horizon limit as
Eq.~\eqref{eq:horizonresponse} gives
\begin{equation}
 \ell_\psi=\sqrt{\frac{3\pi}{32}}
 \int d^3x\,d^3x'\,W_\psi(\mathbf x)W_\psi(\mathbf x')
 |\mathbf x-\mathbf x'|.
 \label{eq:nodalresponse}
\end{equation}
For positive Gaussian envelopes this reduces to $s_\psi$.
A signed interference average need not be determined by the
RMS separation of the squared wave function.
At $w=1$ fm it is 0.80 fm, compared with the Gaussian value
0.67 fm; in the broad-source limit the corresponding values
are 0.85 and 0.90 fm. The no-field nodal projection vanishes
at $w=r_J=0.50$ fm, where its normalized phase response is
singular. Near that cancellation a weak pulse can even increase
the projection by changing destructive interference.

Direct radial normalization, orthogonality and horizon integrals
check Eq.~\eqref{eq:nodalresponse} in the analysis files.
The source-dependent difference cannot be represented by the
same RMS replacement at every $w$.
This calculation therefore does not validate a small nodal
correction: it establishes that the reference shape is a
material assumption in converting the measured slope to lengths.
The main figures and bounds apply to positive Gaussian envelopes,
not to arbitrary nodal or molecular wave functions.

\FloatBarrier
\Needspace{10\baselineskip}
\section*{Acknowledgments}
The author is grateful for the support provided by the
Institute of Physics, Academia Sinica, and the Department
of Physics, National Cheng Kung University. This work was
supported in part by the National Science and Technology
Council of Taiwan.

The author used AI-based language tools for editorial assistance,
including readability, wording, and organization. All scientific
content, calculations, figures, interpretations, and conclusions
were developed, checked, and approved by the author.

\section*{Data availability}
The numerical experimental data used in this study are taken from
the published LHCb measurements cited in the text. Where numerical
tables are unavailable, values are reconstructed from published
figures as described in the corresponding sections. The analysis
code and derived numerical results are available from the author
upon reasonable request.

\FloatBarrier
\setlength{\bibsep}{0pt}


\begin{thebibliography}{99}
\bibitem{Belle} S. K. Choi et al. (Belle Collaboration),
Observation of a narrow charmonium-like state in exclusive
$B^\pm\to K^\pm\pi^+\pi^-J/\psi$ decays,
\href{https://doi.org/10.1103/PhysRevLett.91.262001}{Phys. Rev. Lett. \textbf{91}, 262001 (2003)},
\href{https://arxiv.org/abs/hep-ex/0309032}{arXiv:hep-ex/0309032}.
\bibitem{Quantum} R. Aaij et al. (LHCb Collaboration),
Determination of the $X(3872)$ meson quantum numbers,
\href{https://doi.org/10.1103/PhysRevLett.110.222001}{Phys. Rev. Lett. \textbf{110}, 222001 (2013)},
\href{https://arxiv.org/abs/1302.6269}{arXiv:1302.6269}.
\bibitem{XYZreview} N. Brambilla et al.,
The $XYZ$ states: experimental and theoretical status and perspectives,
\href{https://doi.org/10.1016/j.physrep.2020.05.001}{Phys. Rep. \textbf{873}, 1 (2020)},
\href{https://arxiv.org/abs/1907.07583}{arXiv:1907.07583}.
\bibitem{Factorization} E. Braaten and M. Kusunoki,
Factorization in the production and decay of the $X(3872)$,
\href{https://doi.org/10.1103/PhysRevD.72.014012}{Phys. Rev. D \textbf{72}, 014012 (2005)},
\href{https://arxiv.org/abs/hep-ph/0506087}{arXiv:hep-ph/0506087}.
\bibitem{Production} P. Artoisenet and E. Braaten,
Production of the $X(3872)$ at the Tevatron and the LHC,
\href{https://doi.org/10.1103/PhysRevD.81.114018}{Phys. Rev. D \textbf{81}, 114018 (2010)},
\href{https://arxiv.org/abs/0911.2016}{arXiv:0911.2016}.
\bibitem{ProductionScales} M. Albaladejo, F.-K. Guo, C. Hanhart,
U.-G. Mei\ss ner, J. Nieves, A. Nogga and Z. Yang,
Note on $X(3872)$ production at hadron colliders and its molecular structure,
\href{https://doi.org/10.1088/1674-1137/41/12/121001}{Chin. Phys. C \textbf{41}, 121001 (2017)},
\href{https://arxiv.org/abs/1709.09101}{arXiv:1709.09101}.
\bibitem{Yang} Y. Yang,
Causal-horizon scaling of quarkonium suppression in strong QCD fields,
Phys. Rev. C \textbf{114}, 034904 (2026),
\href{https://doi.org/10.1103/53rr-wr79}{doi:10.1103/53rr-wr79},
\href{https://arxiv.org/abs/2603.24623}{arXiv:2603.24623}.
\bibitem{Xproduction} R. Aaij et al. (LHCb Collaboration),
Measurement of $\chi_{c1}(3872)$ production in proton-proton collisions
at $\sqrt{s}=8$ and 13 TeV,
\href{https://doi.org/10.1007/JHEP01(2022)131}{J. High Energy Phys. \textbf{01} (2022) 131},
\href{https://arxiv.org/abs/2109.07360}{arXiv:2109.07360}.
\bibitem{PsiProduction} R. Aaij et al. (LHCb Collaboration),
Measurement of $\psi(2S)$ production cross-sections in proton-proton
collisions at 7 and 13 TeV,
\href{https://doi.org/10.1140/epjc/s10052-020-7638-y}{Eur. Phys. J. C \textbf{80}, 485 (2020)},
\href{https://arxiv.org/abs/1908.03099}{arXiv:1908.03099}.
\bibitem{JpsiProduction} R. Aaij et al. (LHCb Collaboration),
Measurement of forward $J/\psi$ production cross-sections in $pp$
collisions at $\sqrt{s}=13$ TeV,
\href{https://doi.org/10.1007/JHEP10(2015)172}{J. High Energy Phys. \textbf{10} (2015) 172},
\href{https://arxiv.org/abs/1509.00771}{arXiv:1509.00771}.
\bibitem{JpsiErratum} R. Aaij et al. (LHCb Collaboration),
Erratum to: Measurement of forward $J/\psi$ production cross-sections
in $pp$ collisions at $\sqrt{s}=13$ TeV,
\href{https://doi.org/10.1007/JHEP05(2017)063}{J. High Energy Phys. \textbf{05} (2017) 063}.
\bibitem{CMSproduction} S. Chatrchyan et al. (CMS Collaboration),
Measurement of the $X(3872)$ production cross section via decays to
$J/\psi\pi^+\pi^-$ in $pp$ collisions at $\sqrt{s}=7$ TeV,
\href{https://doi.org/10.1007/JHEP04(2013)154}{J. High Energy Phys. \textbf{04} (2013) 154},
\href{https://arxiv.org/abs/1302.3968}{arXiv:1302.3968}.
\bibitem{ATLASproduction} M. Aaboud et al. (ATLAS Collaboration),
Measurements of $\psi(2S)$ and $X(3872)\to J/\psi\pi^+\pi^-$ production
in $pp$ collisions at $\sqrt{s}=8$ TeV with the ATLAS detector,
\href{https://doi.org/10.1007/JHEP01(2017)117}{J. High Energy Phys. \textbf{01} (2017) 117},
\href{https://arxiv.org/abs/1610.09303}{arXiv:1610.09303}.
\bibitem{Xactivity} R. Aaij et al. (LHCb Collaboration),
Observation of multiplicity-dependent prompt $\chi_{c1}(3872)$ and
$\psi(2S)$ production in $pp$ collisions,
\href{https://doi.org/10.1103/PhysRevLett.126.092001}{Phys. Rev. Lett. \textbf{126}, 092001 (2021)},
\href{https://arxiv.org/abs/2009.06619}{arXiv:2009.06619}.
\bibitem{Compact} A. Esposito, E. G. Ferreiro, A. Pilloni,
A. D. Polosa and C. A. Salgado,
The nature of $X(3872)$ from high-multiplicity $pp$ collisions,
\href{https://doi.org/10.1140/epjc/s10052-021-09425-w}{Eur. Phys. J. C \textbf{81}, 669 (2021)},
\href{https://arxiv.org/abs/2006.15044}{arXiv:2006.15044}.
\bibitem{Molecule} E. Braaten, L. He, K. Ingles and J. Jiang,
Production of $X(3872)$ at high multiplicity,
\href{https://doi.org/10.1103/PhysRevD.103.L071901}{Phys. Rev. D \textbf{103}, L071901 (2021)},
\href{https://arxiv.org/abs/2012.13499}{arXiv:2012.13499}.
\bibitem{OQS} N. Brambilla, M. A. Escobedo, J. Soto and A. Vairo,
Quarkonium suppression in heavy-ion collisions: An open quantum system approach,
\href{https://doi.org/10.1103/PhysRevD.96.034021}{Phys. Rev. D \textbf{96}, 034021 (2017)},
\href{https://arxiv.org/abs/1612.07248}{arXiv:1612.07248}.
\bibitem{Stochastic} Y. Akamatsu and A. Rothkopf,
Stochastic potential and quantum decoherence of heavy quarkonium in the
quark-gluon plasma,
\href{https://doi.org/10.1103/PhysRevD.85.105011}{Phys. Rev. D \textbf{85}, 105011 (2012)},
\href{https://arxiv.org/abs/1110.1203}{arXiv:1110.1203}.
\bibitem{SpatialNoise} S. Kajimoto, Y. Akamatsu, M. Asakawa and A. Rothkopf,
Dynamical dissociation of quarkonia by wave function decoherence,
\href{https://doi.org/10.1103/PhysRevD.97.014003}{Phys. Rev. D \textbf{97}, 014003 (2018)},
\href{https://arxiv.org/abs/1705.03365}{arXiv:1705.03365}.
\bibitem{CharmActivity} R. Aaij et al. (LHCb Collaboration),
Multiplicity dependence of $\sigma_{\psi(2S)}/\sigma_{J/\psi}$ in $pp$
collisions at $\sqrt{s}=13$ TeV,
\href{https://doi.org/10.1007/JHEP05(2024)243}{J. High Energy Phys. \textbf{05} (2024) 243},
\href{https://arxiv.org/abs/2312.15201}{arXiv:2312.15201}.
\bibitem{Satz} H. Satz,
Colour deconfinement and quarkonium binding,
\href{https://doi.org/10.1088/0954-3899/32/3/R01}{J. Phys. G \textbf{32}, R25 (2006)},
\href{https://arxiv.org/abs/hep-ph/0512217}{arXiv:hep-ph/0512217}.
\bibitem{CoupledChannel} M. Cardoso, G. Rupp and E. van Beveren,
Unquenched quark-model calculation of $X(3872)$ electromagnetic decays,
\href{https://doi.org/10.1140/epjc/s10052-014-3254-z}{Eur. Phys. J. C \textbf{75}, 26 (2015)},
\href{https://arxiv.org/abs/1411.1654}{arXiv:1411.1654}.
\bibitem{DMesonRadius} K. U. Can, G. Erkol, M. Oka, A. Ozpineci and T. T. Takahashi,
Vector and axial-vector couplings of $D$ and $D^*$ mesons in 2+1 flavor Lattice QCD,
\href{https://doi.org/10.1016/j.physletb.2012.12.050}{Phys. Lett. B \textbf{719}, 103 (2013)},
\href{https://arxiv.org/abs/1210.0869}{arXiv:1210.0869}.
\bibitem{Lineshape} R. Aaij et al. (LHCb Collaboration),
Study of the lineshape of the $\chi_{c1}(3872)$ state,
\href{https://doi.org/10.1103/PhysRevD.102.092005}{Phys. Rev. D \textbf{102}, 092005 (2020)},
\href{https://arxiv.org/abs/2005.13419}{arXiv:2005.13419}.
\bibitem{RecentPole} T. Ji, X.-K. Dong, F.-K. Guo, C. Hanhart and U.-G. Mei\ss ner,
Precise determination of the properties of $X(3872)$ and of its isovector partner $W_{c1}$,
\href{https://doi.org/10.1016/j.scib.2026.07.071}{Science Bulletin \textbf{71}, 4416 (2026)},
\href{https://arxiv.org/abs/2502.04458}{arXiv:2502.04458}.
\bibitem{WeakBinding} M. Albaladejo and J. Nieves,
Compositeness of S-wave weakly-bound states from next-to-leading order Weinberg's relations,
\href{https://doi.org/10.1140/epjc/s10052-022-10695-1}{Eur. Phys. J. C \textbf{82}, 724 (2022)},
\href{https://arxiv.org/abs/2203.04864}{arXiv:2203.04864}.
\bibitem{CoreProduction} C. Meng, H. Han and K.-T. Chao,
$X(3872)$ and its production at hadron colliders,
\href{https://doi.org/10.1103/PhysRevD.96.074014}{Phys. Rev. D \textbf{96}, 074014 (2017)},
\href{https://arxiv.org/abs/1304.6710}{arXiv:1304.6710}.
\end{thebibliography}
\end{document}